\documentclass[global,twocolumn]{svjour}
\usepackage{graphicx}
\usepackage{epstopdf}

\begin{document}

\title{A simple 2~W continuous-wave laser system for trapping ultracold metastable helium atoms at the 319.8~nm magic wavelength}

\author{R.J. Rengelink \and R.P.M.J.W. Notermans \and W. Vassen*}

\institute{LaserLaB, Department of Physics and Astronomy, Vrije Universiteit Amsterdam,\\ De Boelelaan 1081, 1081 HV Amsterdam, The Netherlands }

\maketitle

\email{$^*$w.vassen@vu.nl} 



\begin{abstract}
High precision spectroscopy on the $2 \ ^3 S \rightarrow 2 \ ^1 S$ transition is possible in ultracold optically trapped helium but the accuracy is limited by the ac-Stark shift induced by the optical dipole trap. To overcome this problem, we have built a trapping laser system at the predicted magic wavelength of 319.8~nm. Our system is based on frequency conversion using commercially available components and produces over 2~W of power at this wavelength. With this system, we show trapping of ultracold atoms, both thermal ($\sim0.2 \ \mathrm{\mu K}$) and in a Bose-Einstein condensate, with a trap lifetime of several seconds, mainly limited by off-resonant scattering.
\end{abstract}



\section{Introduction}
The helium atom has proven to be a productive testing ground for fundamental physics. Frequency metrology has been employed as a sensitive test of QED calculations, both from the ground state~\cite{Bergeson98,Kandula10}, and from the long lived (lifetime $\sim 8000 \ \mathrm{s}$) metastable $2 \ ^3 S$ state ($\mathrm{He^*}$)~\cite{Dorrer97,Luo13,Notermans14}. Another interesting target for spectroscopy is to probe the influence on level energies of the finite size of the nucleus. By comparing accurate atomic structure calculations~\cite{Pach15} to high precision isotope shift measurements, nuclear charge radii relative to the (accurately known~\cite{Sick08}) $\mathrm{^4 He}$ nucleus can be extracted. This method was employed to determine charge radii of the halo nuclei $\mathrm{^6 He}$ and $\mathrm{^8 He}$~\cite{Mueller07} but also to measure the $\mathrm{^4He}$-$\mathrm{^3He}$ differential nuclear charge radius~\cite{Cancio12,Rooij11}. These measurements are relevant to current investigations into the so-called ``proton radius puzzle'' which arose when a similar measurement of the proton radius in $\mathrm{\mu H}$ found a $7 \sigma$ discrepancy with the 2010 CODATA value~\cite{Antog13}. Current efforts investigating the nuclear charge radii of $\mathrm{\mu ^3 He^+}$ and $\mathrm{\mu ^4 He^+}$ are projected to reach an experimental uncertainty at the sub-attometer (am) level~\cite{Nebel12}. Determinations of the $\mathrm{^4He}$-$\mathrm{^3He}$ differential nuclear charge radius with comparable accuracy in electronic systems provide a valuable cross-check for these measurements. 

\begin{figure*}[htbp]
\centering \includegraphics[height=8cm]{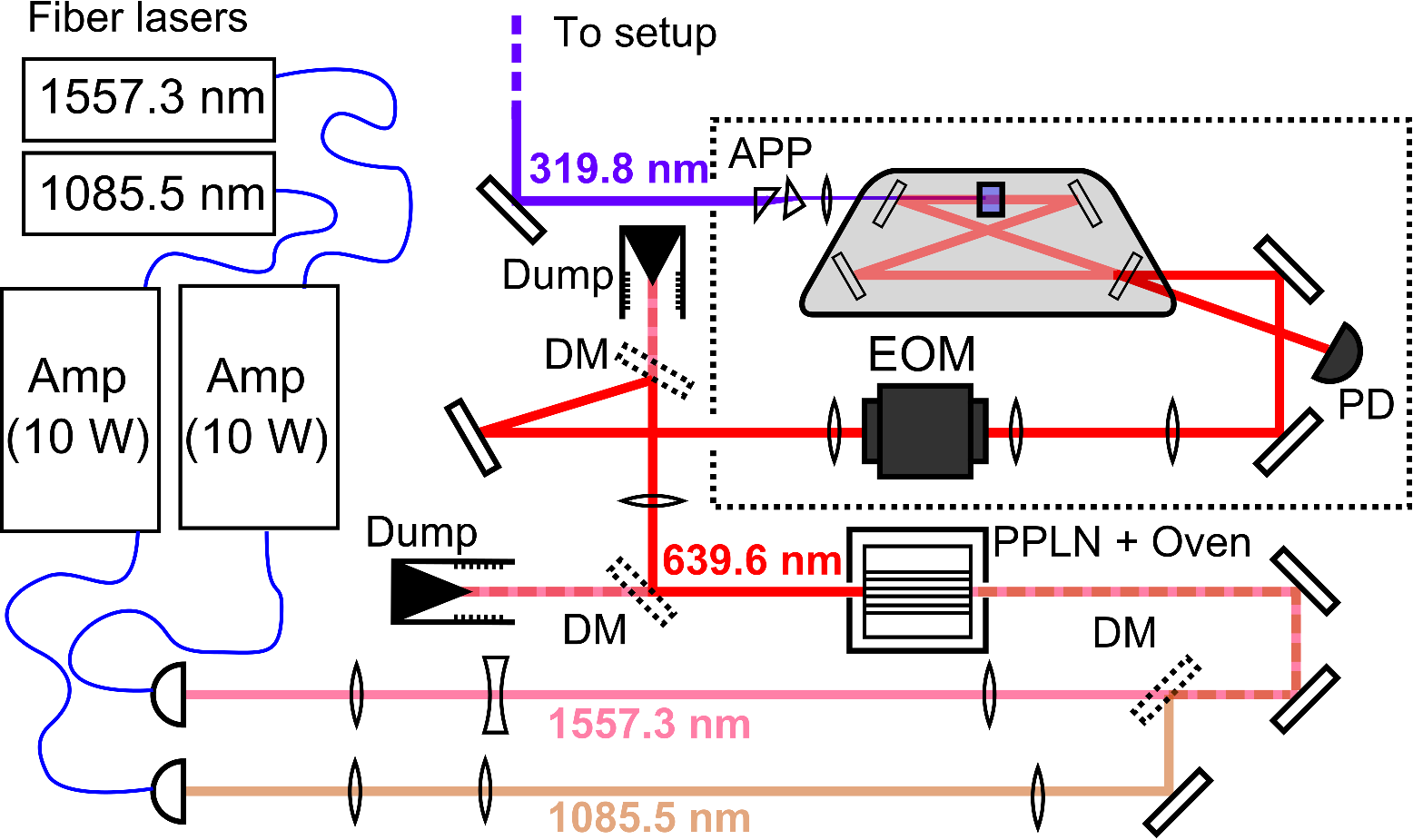}
\caption{Schematic view of the UV laser system. Two infrared fiber lasers at 1085.5~nm and 1557.3~nm seed two 10~W amplifiers. The infrared beams are independently focussed, and overlapped on a dichroic mirror (DM). The combined beam passes through a temperature stabilized PPLN crystal. More dichroic mirrors filter residual infrared light from the SFG beam at 639.6~nm. This beam is mode-matched and phase modulated by a 20~MHz electro-optic modulator (EOM) and the cavity reflection is monitored using a photodiode (PD) to allow Pound-Drever-Hall locking of the cavity. The final UV output beam is then collimated and ellipticity is compensated by an anamorphic prism pair (APP).}\label{fig:setup}
\end{figure*}

Currently the two most accurate measurements of the $\mathrm{^4He}$-$\mathrm{^3He}$ differential nuclear charge radius have achieved accuracies of 3~\cite{Cancio12} and 11 $\mathrm{am^2}$~\cite{Rooij11}, roughly an order of magnitude less precise than the projection of the $\mathrm{\mu He}$ experiment, but disagree by $4 \mathrm{\sigma}$. The former experiment resolved the $2 \ ^3 S \rightarrow 2 \ ^3 P$ transition to within one-thousandth of the 1.6 MHz natural linewidth and is not expected to be improved upon in the near future. The latter experiment was performed on the doubly forbidden $2 \ ^3 S \rightarrow 2 \ ^1 S$ transition whose 8 Hz natural linewidth is not a limiting factor but has a very low excitation rate and thus requires a long interaction time. To achieve this $\mathrm{He^*}$ atoms were cooled to quantum degeneracy (a Bose-Einstein condensate (BEC) of $\mathrm{^4He^*}$, and in a degenerate Fermi gas of $\mathrm{^3 He^*}$) and trapped in an optical dipole trap (ODT). The accuracy of this experiment was limited by experimental effects, mainly the ac-Stark shift induced by the ODT.

This problem is also encountered in optical lattice clocks where it is solved by employing so-called magic wavelength traps~\cite{Ye08}. In a magic wavelength trap, the wavelength of the trapping laser is chosen such that the upper and lower state polarizability are exactly the same, cancelling out the differential ac-Stark shift. In helium a high-precision calculation of the ac-polarizability of the $2 \ ^3S$ level was recently reported~\cite{Mitroy13}, and we ourselves have made more approximate calculations on both the $2 \ ^3 S$ and  $2 \ ^1 S$ levels~\cite{Noter14}. Both works predict the polarizability of the $2 \ ^3 S$ level to vanish at around 413~nm (a so-called tune-out wavelength) which was later confirmed experimentally~\cite{Henson15}. Our calculations also predict a number of magic wavelengths for the $ 2 \ ^3 S \rightarrow 2 \ ^1 S$ transition. The most promising from the perspective of trapping is located at 319.815~nm for $\mathrm{^4 He}$ and 319.830~nm for $\mathrm{^3 He}$.

Trapping atoms at this wavelength is not straightforward. First of all, to achieve a trap depth comparable to~\cite{Rooij11}, where an infrared ODT at 1557~nm was used, an optical power of approximately 1~W is required. Such powers are not readily available at ultraviolet wavelengths. Secondly, the lifetime of atoms trapped at this magic wavelength is intrinsically limited by two mechanisms: off-resonant excitation to the nearby $2 \ ^3 S \rightarrow 4 \ ^3 P$ transition at 318.9~nm, and two-photon ionization. The total loss rate from these processes should not exceed about 1 $s ^{-1}$, to allow sufficient probe time for spectroscopy.

Considerable progress has been made in the production of laser light in the wavelength range near 320~nm, primarily for the purpose of laser cooling $\mathrm{Be^+}$ ions~\cite{Vasil11,Cozijn13,Wilson11,Lo14,Hankin14}. High power (several hundreds~mW) was generated by sum frequency mixing and subsequent frequency doubling of two fiber lasers, retaining most of their high spatial and spectral mode quality~\cite{Wilson11,Lo14,Hankin14}. Production of up to 2~W was demonstrated with such a system~\cite{Lo14}. Constrained to commercial Er and Yb doped fiber amplifiers, this scheme allows the production of high power continuous wave laser light over a range of 310-325~nm.

In section~\ref{sec:lasersys}, we demonstrate a laser system built out of commercially available components producing over 2~W at 319.8~nm based on a modification of this scheme. In section~\ref{sec:trapping} we show that our source can be used to trap helium atoms with an acceptable lifetime of a few seconds, such that spectroscopy on the $2 \ ^3 S \rightarrow 2 \ ^1 S$ transition can be performed.
\section{Laser System}
\label{sec:lasersys}
The system can be divided in a sum frequency generation (SFG) part, which generates 639.6~nm light from two infrared lasers, and a second harmonic generation (SHG) part which frequency doubles the SFG light to 319.8~nm. In the following, we will first give a full overview of the optical system before discussing the results and performance of the SFG and SHG parts separately.
\subsection{Overview}
\label{sec:overview}
Figure~\ref{fig:setup} shows a schematic overview of the optical setup. The setup is relatively compact with all of the components, except for the lasers and the control electronics, mounted on a single 1000$\times$500 $\mathrm{mm^2}$ optical breadboard. The system starts with two fiber lasers (NKT photonics Koheras Adjustik E15 and Y10) with center wavelengths of 1557.28~nm and 1085.45~nm. The thermal tuning ranges are 1000~pm and 700~pm respectively, covering a spectral range much larger than the uncertainty in the calculated magic wavelength. The lasers seed two 10~W fiber amplifiers (NuFern NUA-1084-PB-0010-C2 and NUA-1550-PB-0010-C2), with isolated polarization maintaining free-space output couplers. The beams are separately focussed to achieve optimal sum frequency generation.

The output beams are to an excellent degree of approximation Gaussian, and can be described completely by two parameters: their (minimum) waist size ($w_0$), which is directly related to their Rayleigh range ($z_R = \pi w_0^2/\lambda$), and the position of their focus. In order to achieve optimal conversion these parameters must be matched both to each other and to the crystal. Achieving this condition is not entirely straightforward because of the coupled nature of the problem.

We first collimate the 1557.3~nm beam with an $f = 300 \ \mathrm{mm}$ lens, and then focus by two lenses with focal distances of -100~mm and 200~mm. The beam waist can now be changed by moving either of the focussing lenses, but doing so will also move the focal point. The 1085.5~nm beam is first passed through a telescope consisting of two lenses with focal distances of 200~mm and 50~mm and is then focussed by an $f = 300 \ \mathrm{mm}$ lens. The focal point can be changed without affecting the waist size by moving the focussing lens. In this way, the foci of the beams are matched by first setting the waist of the 1085.5~nm beam to the desired focussing, secondly matching the 1557.3~nm beam waist to it, and finally overlapping the focal point of the 1085.5~nm beam with that of the 1557.3~nm beam.

With fixed beam parameters, the infrared beams are overlapped on a dichroic mirror and passed through a 40~mm MgO doped periodically poled lithium niobate (PPLN) crystal (Covesion) with a poling period of $12.1 \ \mathrm{\mu m}$. The crystal is mounted in an oven and temperature stabilized at $ \sim 90^\circ$C. The output beam from the crystal contains both the sum frequency and residual infrared light. This residual light is filtered from the beam by two dichroic mirrors and the SFG beam is collimated by a $f = 250 \ \mathrm{mm}$ lens to a waist of $\sim 1 \ \mathrm{mm}$. 

The light is then coupled (free-space) into a commercial frequency doubling system (Toptica SHG~pro) where it is mode-matched to the cavity and passed through an electro-optical modulator (EOM). The EOM modulates 20~MHz sidebands on the laser carrier frequency to allow Pound-Drever-Hall locking of the cavity. The doubling cavity is similar to \cite{Koel05}, the main differences being the locking scheme (Pound-Drever-Hall instead of H\"ansch-Couillaud) and the crystal (AR-coated rather than Brewster cut). The UV output is collimated and passed through an anamorphic prism pair to reduce beam ellipticity. Based on the specifications of the seed lasers and amplifiers the spectral linewidths of the infrared beams should be of the order of several tens of kHz. Because the nonlinear conversion steps do not significantly add to the fractional linewidth the final UV-output is expected to have a linewidth of $\sim 100\ \mathrm{kHz}$, which is small compared to the scale at which the polarizability changes~\cite{Noter14}.
\subsection{Sum Frequency Generation}
\begin{figure}[htbp]
\centering
\includegraphics[width=\columnwidth]{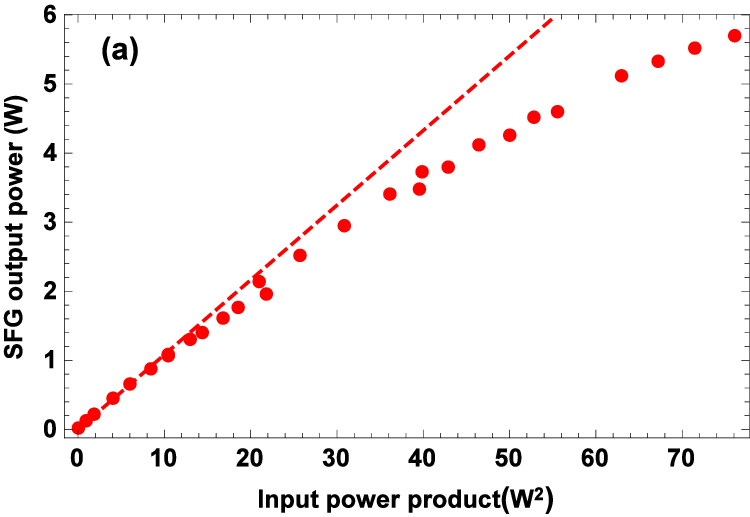}
\includegraphics[width=\columnwidth]{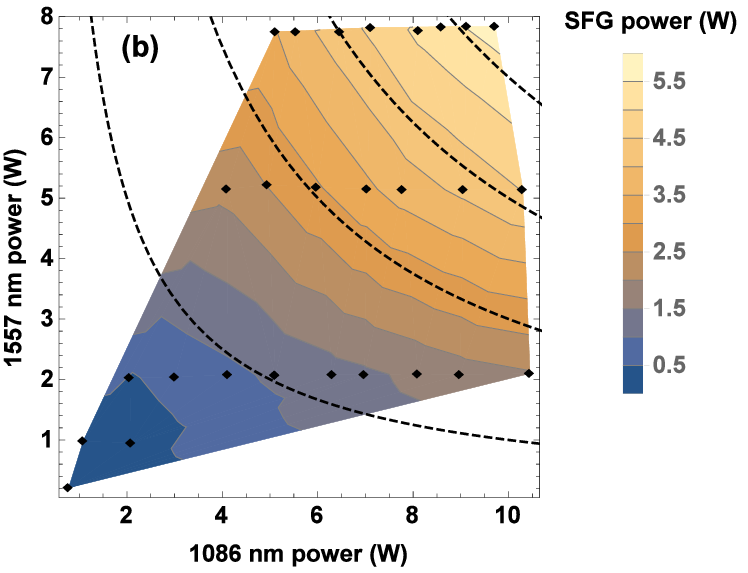}
\caption{Results of sum frequency generation. (a) Sum frequency power as a function of the input power product (IPP) of the infrared lasers. The dashed line is a linear fit at low input powers (slope $0.108(1) \ \mathrm{W^{-1}}$). (b) Contourplot of the sum frequency power as a function of both input powers, based on a linear interpolation of the same dataset as (a). Black diamonds indicate measured datapoints, the dashed lines indicate contours of constant IPP.}\label{fig:SFGconv}
\end{figure}
The purpose of the SFG stage is to convert the available infrared laser light into useful SFG light with high efficiency. To achieve this it is necessary to focus the input beams tightly so that a high peak intensity is reached, but not so tightly that the beams diverge too quickly before they reach the end of the crystal.  As described by Boyd and Kleinman~\cite{Boyd68}, this process can be optimized with respect to the dimensionless focussing parameter $\xi = l / 2 z_R$, which is the ratio of the crystal length $l$ to the confocal parameter of the beam (twice the Rayleigh length $z_R$). Although optimal at $\xi \approx 2.84$, the efficiency varies quite slowly so that at confocal focussing ($\xi = 1$) it is still approximately 80\% of its maximum value. The system produces more power than required and the confocal condition is chosen because of practical considerations such as the available path length and the size of the entrance surface of the crystal.

The infrared beams were set to the confocal condition as described in section~\ref{sec:overview}. The waist sizes are measured to be $56(1) \ \mathrm{\mu m}$ ($z_R = 8.9(3) \ \mathrm{mm}$) for the 1085.5~nm beam and $63(1) \ \mathrm{\mu m}$ ($z_R = 7.9(3) \ \mathrm{mm}$) for the 1557.3~nm beam, with their waist positions located within 1~mm of each other. By correcting for the refractive index of the crystals, which can be calculated based on known Sellmeier coefficients~\cite{Gayer08}, the focussing parameter inside the crystal is found to be $\xi = 1.03$ for the 1085.5~nm beam and $\xi = 1.16$ for the 1557.3~nm beam.

Figure~\ref{fig:SFGconv}a shows the converted power as a function of the product of the input powers (input power product, IPP). When different combinations of input powers with equal IPP are used they produce almost the same SFG output power. This can be seen more clearly in a contour plot of the output power as a function of input powers (figure~\ref{fig:SFGconv}b, based on a linear interpolation of the same dataset) where contours of constant IPP follow constant output power. From this we conclude that the total converted power is a function of IPP only, and does not depend on the exact composition of infrared powers.

\begin{figure}[tbp!]
\centering\includegraphics[width=\columnwidth]{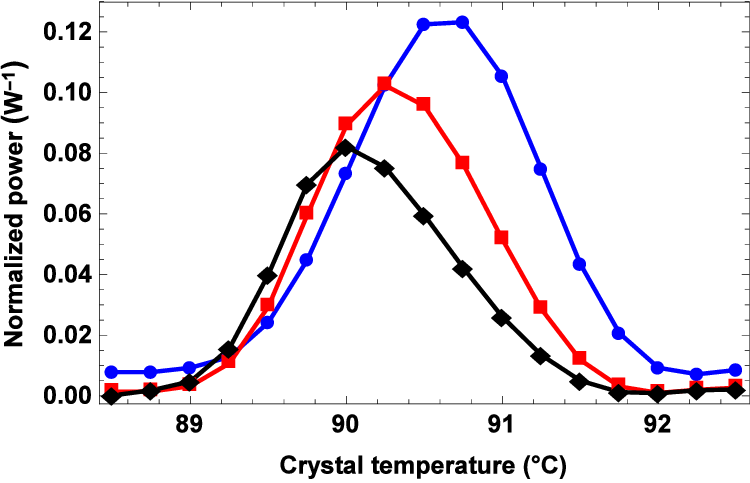}
\caption{Sum frequency production as a function of crystal temperature, normalized for measured input power product. The plot shows input power products of $1.4 \ \mathrm{W^2}$ (blue circles), $35.7 \ \mathrm{W^2}$ (red squares), and  $78.0 \ \mathrm{W^2}$ (black diamonds). At higher input powers the temperature graph becomes slanted towards lower temperature, indicating thermal effects.}\label{fig:SFGtemp}
\end{figure}

In order to achieve the power conversion plotted in Figs.~\ref{fig:SFGconv}a and \ref{fig:SFGconv}b, it is necessary to optimize for crystal temperature at each input power product. The reason for this is shown in Figure~\ref{fig:SFGtemp}, which shows the conversion efficiency as a function of crystal temperature at different input powers. As the input power becomes higher, the optimal temperature shifts to lower temperature and the crystal temperature needs to be adjusted to achieve maximum conversion. A possible explanation for this behaviour is that light is absorbed in the center of the crystal and heats it locally. This causes a slightly elevated temperature, and consequently imperfect phase matching at the center of the crystal (where conversion occurs) compared to the crystal edge (with respect to which the temperature is controlled). When the crystal temperature is set to a slightly lower temperature, this effect is compensated. 

At low input powers the SFG output power scales linearly with a slope of $0.108(1) \ \mathrm{W^{-1}}$, comparable to other experiments using a similar crystal~\cite{Wilson11,Lo14}. At higher output powers a deviation from the linear behaviour is seen. This may be a left-over thermal effect, or it may be that, because of the high conversion efficiency, pump depletion needs to be taken into account. In the case of a thermal effect the spatial output mode may be distorted but this is not observed. At an IPP of $80 \ \mathrm{W^2}$, consisting of 8~W at 1557.3~nm and 10~W at 1085.5~nm, a maximum output power of almost 6~W of SFG light is produced, which corresponds to a conversion efficiency of 33\%.
\subsection{Second Harmonic Generation}
The generated SFG light is coupled in free space to the Toptica SHG pro system. Using a commercial frequency doubling system has advantages, but the disadvantage is that some system parameters are not disclosed. We will therefore describe the system as a whole with reported in- and output powers measured before and after the full system.

\begin{figure}[hb!]
\centering\includegraphics[width=\columnwidth]{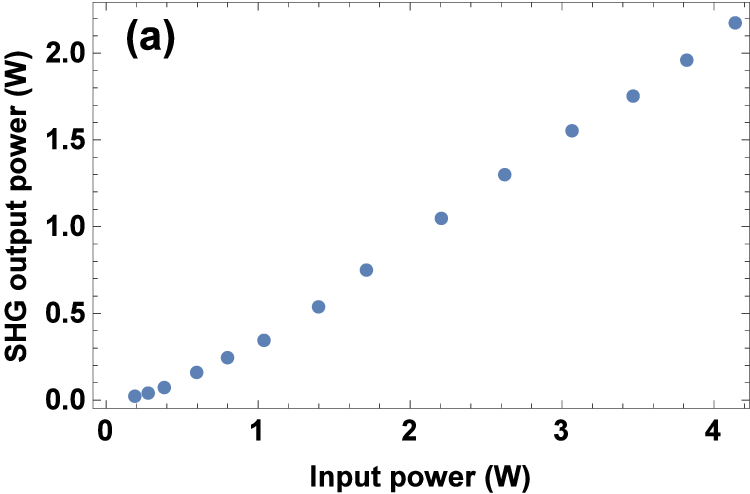}
\includegraphics[width=\columnwidth]{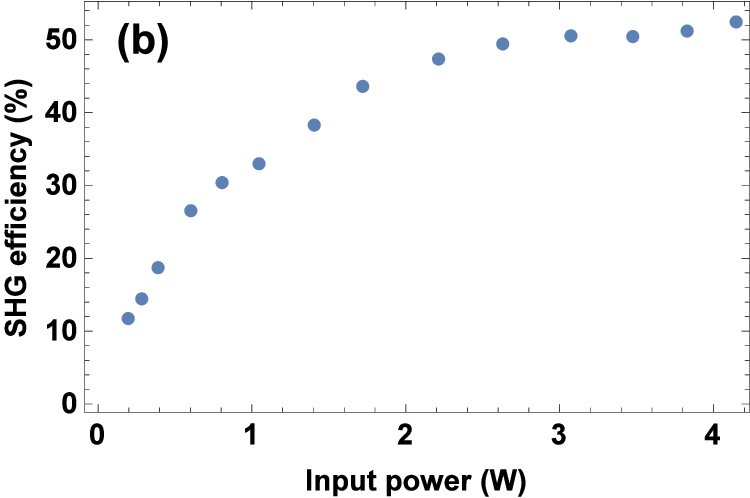}
\caption{Results from the SHG system. (a) UV output power (319.8 nm) as a function of total input power (639.6 nm) going into the full system. The output increases quadratically at low input powers, but saturates to a linear asymptote at higher powers. (b) System conversion efficiency (ratio of total input and output power) as a function of input power. The efficiency saturates at $\sim$50\%.}\label{fig:SHGconv}
\end{figure}

Figure~\ref{fig:SHGconv} shows the system output power and conversion efficiency of the SHG section. At an input power of $\sim$2 W the conversion efficiency saturates at about 50\% and at an input power of 4~W a maximum output power of more than 2~W of UV light was achieved. At this point the cavity coupling efficiency is just over 80\%. This behaviour is qualitatively similar to what is observed in other SHG systems~\cite{Koel05,Targat05}. Although more SFG input power is available, we choose to remain below 4~W input power to prevent (UV-induced) damage to the cavity optics at these powers~\cite{Sim14}. Peak-to-peak UV intensity fluctuations of 4\%  were measured over a 4~hour period. This is within the specifications for intensity stability of the fiber amplifiers. No evidence of degradation of the crystal or cavity optics has been observed after more than a year of operation although periodic realignments of the cavity mirrors were required.

The beam profile of the cavity output is still Gaussian ($M^2 < 1.2$), but shows ellipticity due to crystal walk-off. Directly from the cavity we observe a beam waist of $\sim0.7 \ \mathrm{mm}$ in the vertical and $\sim 0.1 \ \mathrm{mm}$ in the horizontal direction. This beam is passed through a collimating lens and an anamorphic prism pair to produce a more circular beam. We measure the final beam waist of $\sim0.21 \times 0.35 \ \mathrm{mm^2}$. Despite the modest ellipticity these beam parameters still allow tight focussing, an essential requirement for optical dipole trapping.

\section{Trapping}
\label{sec:trapping}
Now that we are able to produce sufficient power at 319.8~nm, we implement the laser system into our existing setup~\cite{Notermans14,Rooij11} to demonstrate trapping and to characterize the trap lifetime. We prepare the beam for trapping by enlarging it with a 1:2 telescope and focus it inside the vacuum chamber with an $f=400 \ \mathrm{mm}$ lens to a waist of $w_0^{(1)} \times w_0^{(2)} = 64.3(1.0)\times 55.6(7) \ \mathrm{\mu m^2}$. The focus positions along the horizontal and vertical axes are found to lie 12(2)~mm apart which is small compared to the Rayleigh lengths (40.0(6)~mm and 30.0(4)~mm). We measure 62\% total transmission of the two windows of the vacuum chamber. While high, these losses are expected from uncoated sapphire vacuum windows~\cite{Eat14}. 

A transmission per window of $T_1 = \sqrt{0.62} \approx 0.78$ is assumed to estimate the power inside the vacuum chamber. Therefore, at a power ($P$) of 1~W and neglecting astigmatism, the beam has peak intensity
\begin{equation}
I_p = \frac{2 T_1 P}{\pi w_0^{(1)} w_0^{(2)}} = 14 \ \mathrm{kW cm^{-2}}.
\end{equation}
Based on the polarizability (calculated in~\cite{Noter14}) this translates to a trap depth of $\sim 6.0 \ \mathrm{\mu K \  W^{-1}}$ for a single beam. The trap depth of our ODT is therefore far below the recoil temperature
\begin{equation}
T_{rec} = \frac{\hbar^2 k^2}{k_B m} = 46.7 \ \mathrm{\mu K},
\end{equation}
where $k = 2 \pi / \lambda$ is the photon wavenumber and $m$ is the atomic mass of helium. We can therefore safely assume that each photon scattering event leads to the loss of the scattered atom. Additionally, the excess energy of the scattered atom can heat the other atoms or even kick more atoms out of the trap.

When two- and three-body losses can be neglected, the total loss rate is a combination of three distinct rates: a background loss rate $R_{bg}$ due to the background pressure inside the chamber, a loss rate $R_{sc}$ due to photon scattering, and a loss rate $R_{ion}$ due to two-photon ionization. These mechanisms scale in different ways with ODT power. The total loss rate is
\begin{equation}
R_{tot} = R_{bg} + R_{sc} I_p + R_{ion} I_p^2.
\label{eq:rates}
\end{equation}

The Rayleigh length of the focussed UV beam is comparable to the $\sim$4 cm  spacing between the vacuum windows. Therefore, in a single beam ODT trapped atoms are able to collide with the windows and leave the trap. To avoid this some means of axial confinement is necessary. We use two different methods to provide this confinement. The first is to add a magnetic field gradient along the beam direction to create a hybrid trap~\cite{Flores15}. With this method the loss rate and trap depth are straightforward to interpret. However, peak densities are not high enough to produce a BEC. The other method is to create a two-colour crossed-beam ODT using the UV beam and additionally a focussed 1557 nm beam which was already in place~\cite{Rooij11}. This trap gives a high enough peak density for a BEC to form. 

The final magic wavelength trap for precision spectroscopy will be of a different geometry however, because both trapping schemes disussed here still introduce systematic shifts of the transition frequency. The most straightforward final trap geometry would be a crossed beam ODT using only the UV laser. In such a trap the atomic density will be similar to what is found in the two-colour trap while the trap depth and scattering rate are comparable to the hybrid trap. Measuring these quantities therefore gives an accurate picture of what can be expected while requiring no modification of the current experimental setup.

\subsection{Hybrid Trap}
\begin{figure}[htbp]
\centering
\includegraphics[width=0.5\textwidth]{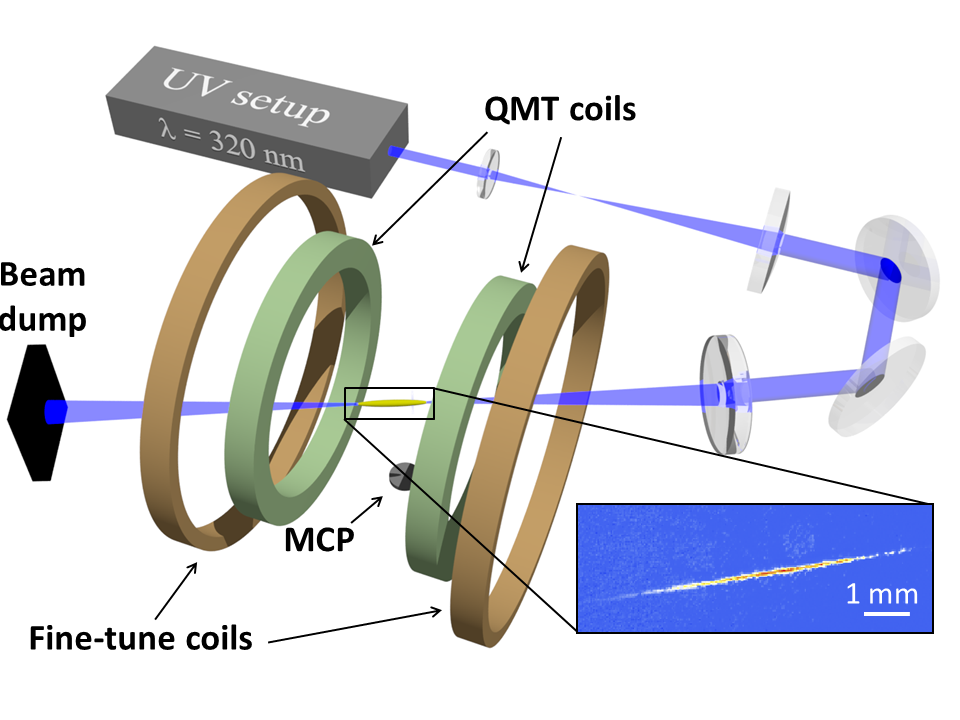}
\caption{Schematic view of the hybrid trap geometry. Atoms are trapped by the combination of the UV-laser beam and the quadrupole magnetic trap (QMT) coils. A set of fine-tune coils allows precise tuning of the trap center. Because of limited optical access the UV laser beam is at an angle of $\sim 9.5^\circ$ with the magnetic field axis. High-resolution detection is done using a micro-channel plate detector (MCP). Inset: in-situ absorption image of atoms in the UV hybrid trap.}\label{fig:Hybridtrap}
\end{figure}
We prepare an ultracold sample in a way previously described~\cite{Rooij11}. A beam of $\mathrm{He^*}$ atoms is generated from a liquid nitrogen cooled dc-discharge, collimated, slowed in a Zeeman slower, and captured in a magneto-optical trap. Here the atoms are cooled to approximately 0.5~mK. Subsequently they are spin-polarized, loaded into a Ioffe-Pritchard type magnetic trap and Doppler cooled to approximately $130 \ \mathrm{\mu K}$. Finally, the atoms are cooled to $\sim0.2 \ \mathrm{\mu K}$ by forced rf evaporative cooling inside the magnetic trap.

The cloud is transferred to the hybrid trap consisting of the UV beam and a quadrupole magnetic trap (QMT) generated by a set of QMT coils in anti-Helmholtz configuration. Figure~\ref{fig:Hybridtrap} shows a schematic of this trap. The quadrupole field has a strong axis gradient of 0.54~Gauss/cm but gravity acts in a direction perpendicular to this axis. In this direction the gradient is only half the magnitude which is well below the leviation gradient of $m g / \mu = 0.351 \ \mathrm{Gauss/cm}$~\cite{Flores15}. Below this gradient gravity is stronger than the confining magnetic force so that atoms are not trapped in the absence of the UV beam. The QMT therefore adds confinement but does not contribute to the trap depth. A homogeneous magnetic field is applied with a set of fine-tune coils to minimize trap oscillations induced by the loading step.

The atoms are detected either by absorption imaging or by a micro-channel plate detector (MCP) located 17~cm below the trap center. MCP time of flight measurements are done as a function of hold time in the hybrid trap. The time of flight signals are fitted with a thermal Bose-Einstein distribution to extract the temperature of the gas as well as the atom number. The decay in atom number is fitted with an exponential (with an oscillating component to account for residual trap oscillations). The first few seconds of the decay are not fitted to neglect two-body loss and thermalization effects.

\begin{figure}[bp]
\centering\includegraphics[width=\columnwidth]{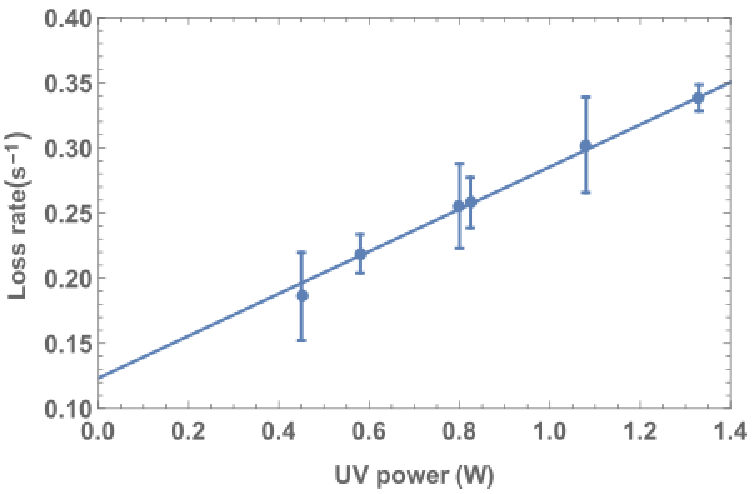}
\caption{Hybrid trap one-body loss rate as a function of UV power. The blue line is a linear fit with a slope of 0.16(2) $\mathrm{s^{-1} W ^{-1}}$, corresponding to off-resonant scattering.}
\label{fig:Lossrate}
\end{figure}

Figure~\ref{fig:Lossrate} shows the loss rate of a thermal gas inside the hybrid trap. The loss rate varies linearly with power, which is consistent with only background collisions and off-resonant scattering; two-photon ionization would depend quadratically on power. A linear fit gives a slope of about 0.16(2) $\mathrm{s^{-1} W^{-1}}$ ($R_{sc} \approx 1.2 \ \mathrm{s^{-1} W^{-1} m^2}$, see equation~\ref{eq:rates}), with a background loss rate of 0.12(2) $\mathrm{s^{-1}}$. This background loss rate is consistent with the loss rate of 0.13(1) $\mathrm{s^{-1}}$ in a hybrid trap of comparable depth using our 1557~nm ODT for which off-resonant scattering is negligible~\cite{Noter14}. 

\begin{figure}[htp]
\centering\includegraphics[width=\columnwidth]{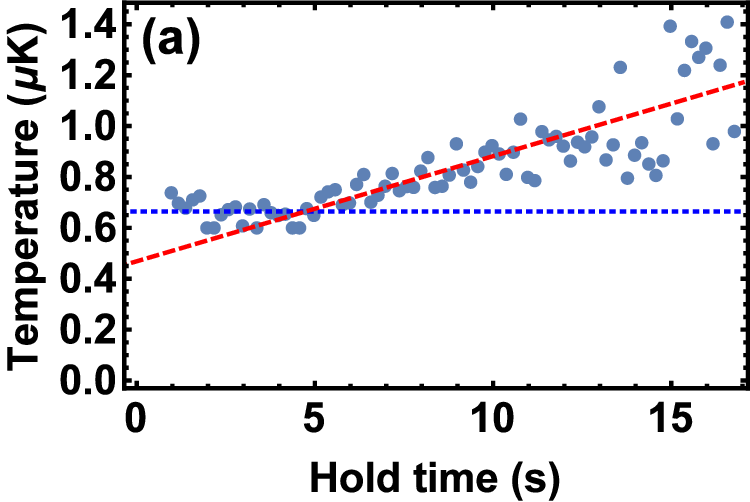}
\includegraphics[width=\columnwidth]{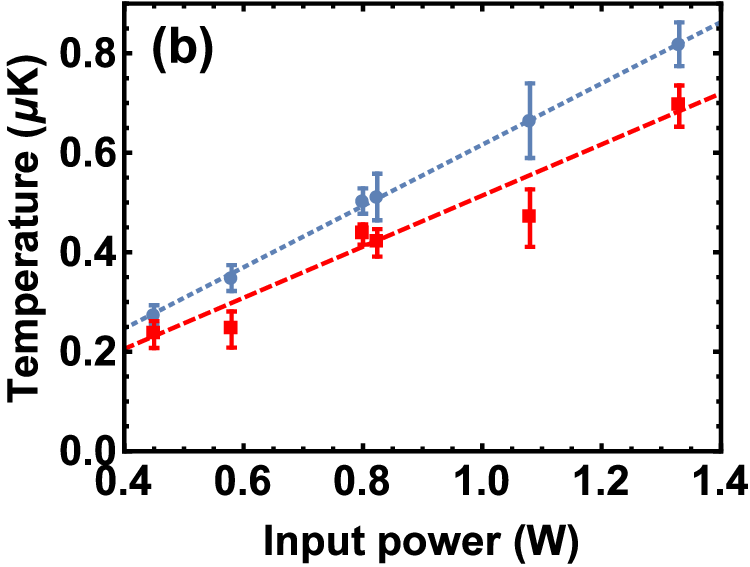}
\caption{(a) Temperature as a function of hold time in the hybrid trap. After some thermalization time the atomic cloud starts to heat as a result of UV absorption. The blue dotted line is a constant fit to the temperature around the minimum to extract the minimum temperature. The red dashed line is a linear fit to the temperature after a few seconds to extract the heating rate. (b) Determination of the trap temperature based on minimum temperature (blue circles), and an extrapolation to zero hold time of the heating rate (red squares). From linear fits to these data an upper (lower) bound is set on the temperature in absence of heating of $0.62(2) \ \mathrm{\mu K \ W^{-1}}$ ($0.51(2) \ \mathrm{\mu K \ W^{-1}}$).}\label{fig:HTtemp}
\end{figure}

Figure~\ref{fig:HTtemp} shows the fitted temperature as a function of hold time in the hybrid trap. This temperature is not constant; after a quick thermalization the temperature starts to increase linearly with time, however, not fast enough to pose a problem. Two generic sources of heating in a dipole trap are intensity noise and beam pointing noise~\cite{Savard97}. The former is not observed at this timescale because it would lead to exponential rather than linear heating. Beam pointing noise is a possible explanation of the observed heating but was never observed to be a problem in our infrared trap which uses a similar geometry. A more plausible heating mechanism is that atoms off-resonantly scattering a photon dump a small portion of their high recoil energy in the atomic cloud, thereby heating the ensemble. In principle it is also possible that a fraction of atoms heated by scattering is able to dump all recoil energy in the cloud and thermalize instead of leaving the trap. This may cause the photon scattering rate that was determined earlier to be an underestimate because a part of all scattering events appear as heating rather than trap loss. To assess the maximum contribution to the scattering rate of this effect we assume the most extreme case in which the rethermalization of recoiling atoms causes all of the observed heating. In this case, the highest observed heating rate of $\dot{T} \approx 0.04 \ \mathrm{\mu K \ s^{-1}}$ corresponds to no more than $\dot{T}/T_{rec} \approx 0.001 \ \mathrm{s^{-1}}$ unaccounted scattering events. This is two orders of magnitude lower than typically observed trap loss and can be safely disregarded.

To give an estimate of the equilibrium temperature inside the trap (in the absence of heating) we take the minimum achieved temperature as an upper bound, and a linear extrapolation of the heating to zero hold time as a lower bound. In this way we extract temperatures of $0.62(2) \ \mathrm{\mu K \ W^{-1}}$ and $0.51(2) \ \mathrm{\mu K \ W^{-1}}$ respectively. This is approximately a factor 10 lower than the calculated trap depth and corresponds to a truncation parameter $\eta=10$ which is typically found in a thermalized trapped gas~\cite{Flores15}.

\subsection{Two-colour Trap}
\begin{figure}[bp!]
\centering\includegraphics[width=\columnwidth]{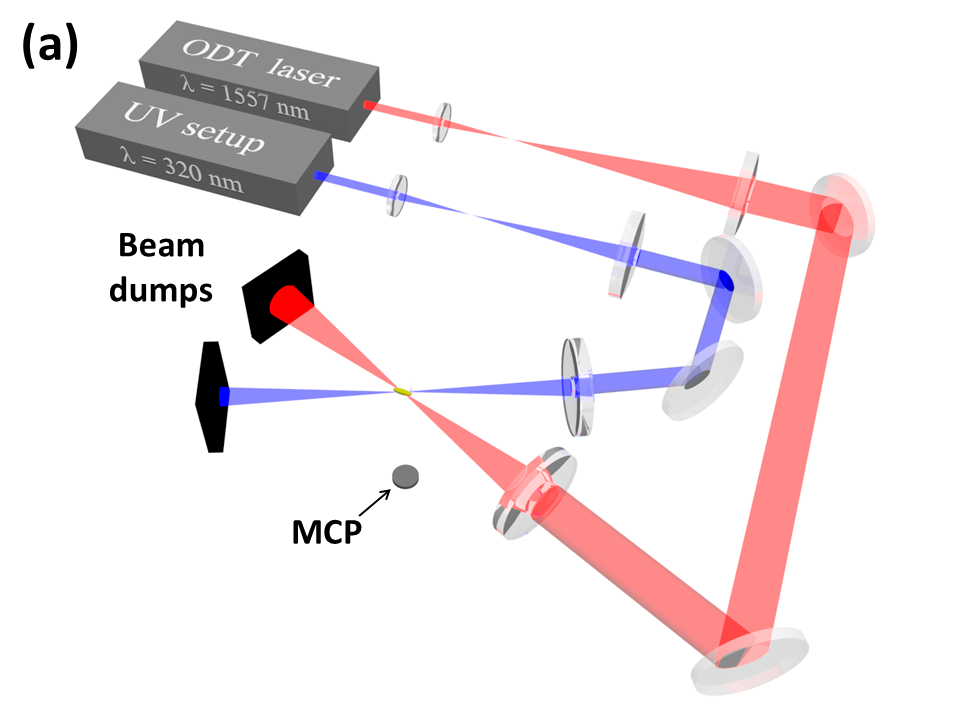} \\\includegraphics[width=\columnwidth]{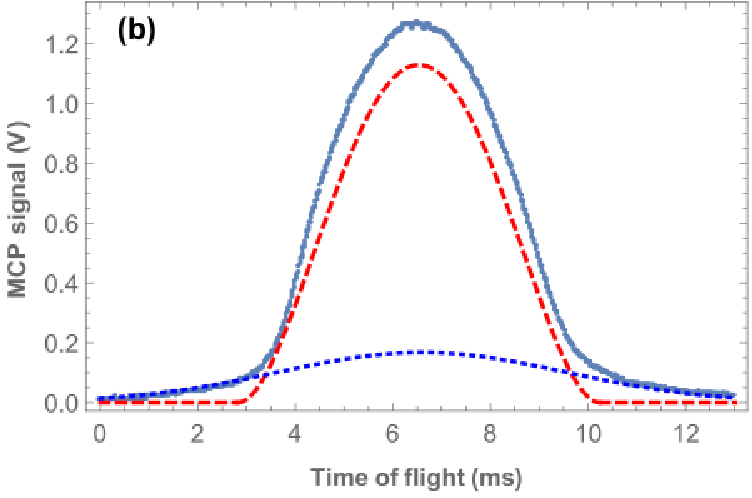}
\caption{(a) Schematic picture of the two-colour ODT setup. The angle between the beams is 19$^\circ$. (b) MCP time-of-flight measurement of a BEC in a two colour optical dipole trap. The signal is fit with a bimodal distribution indicating a BEC (red dashed line) and a thermal fraction (blue dotted line).}\label{fig:TwocolourODT}
\end{figure}

Because of the low confinement provided by the hybrid trap, no BEC was observed. To provide enough confinement to observe BEC we switch to a two-colour trap consisting of the UV beam and an IR beam (focussed to a waist of $85 \ \mathrm{\mu m}$) which cross at an angle of $19^\circ$. Figure~\ref{fig:TwocolourODT}a shows a schematic of the setup. Figure~\ref{fig:TwocolourODT}b shows the corresponding time-of flight signal on the MCP, fit with a bimodal distribution. Superposed on the thermal distribution is a clear inverted parabola Thomas-Fermi profile demonstrating Bose-Einstein condensation. For this BEC a one-body lifetime is observed of $\sim 4 s$.
\section{Conclusion}
In order to perform magic wavelength trapping of metastable helium atoms we have realized a laser system which produces over 2~W at 319.8~nm, with not yet an indication of a reduction of SHG efficiency at higher pump power. The setup is built from commercially available fiber lasers, amplifiers and SFG/SHG components. Similar performance should be possible in a spectral range of 310-325~nm with only minor changes in the required components (mainly limited by available wavelength ranges of the amplifiers). The produced UV light is used to trap an ultracold ($\sim 0.2 \ \mathrm{\mu K}$) thermal gas in a hybrid trap and a BEC in a two colour ODT. Trap losses are found to be mainly due to off-resonant scattering with a rate of 0.16(2) $\mathrm{s^{-1} \ W^{-1}}$. 

With this system we can make a sufficiently deep dipole trap in the UV while keeping the intrinsic losses at an acceptable level such that spectroscopy is possible. This opens the door to a full magic wavelength ODT and a more precise measurement of the $2 \ ^3 S \rightarrow 2 \ ^1 S$ transition frequency.

\begin{acknowledgement}
We would like to thank Christian Rahlff of Covesion for useful comments about the PPLN crystal, Steven Knoop for useful discussions about hybrid trapping and Rob Kortekaas for technical support. This work is part of the research programme of the Foundation for Fundamental Research on Matter (FOM), which is part of the Netherlands Organisation for Scientific Research (NWO).
\end{acknowledgement}

\end{document}